\documentclass[12pt,preprint]{aastex}
\usepackage{amsmath}
\usepackage{graphicx}
\usepackage{epsfig}
\usepackage{url}
\usepackage{indentfirst}
\begin{document}

\title{Observation and Simulation of the Variable Gamma-ray Emission from PSR~J2021+4026}
\author{Ng, C. W.\altaffilmark{1}, Takata, J.\altaffilmark{2} \and Cheng, K. S.\altaffilmark{1}}
\email{rubyngcw@connect.hku.hk, takata@hust.edu.cn, hrspksc@hku.hk}
\altaffiltext{1}{Department of Physics, The University of Hong Kong, Pokfulam Road, Hong Kong}
\altaffiltext{2}{School of physics, Huazhong University of Science and Technology, Wuhan 430074, China}

\begin{abstract}
  Pulsars are rapidly spinning and highly magnetized neutron stars, with highly stable
  rotational period and  gradual spin-down over a long timescale due to the loss of radiation.
  Glitches refer to the events that suddenly increase the rotational speed of a pulsar.
  The exact causes of glitches and the resulting processes are not fully understood.
  It is generally believed that couplings between the normal matter and the superfluid
  components, and the starquakes, are the common causes of glitches. In this study,
  one famous glitching pulsar, PSR~J2021+4026, is investigated. PSR~J2021+4026 is the first variable gamma-ray pulsar observed by \textit{Fermi}. From the gamma-ray observations, it is found that the pulsar experienced a significant flux drop, an increase in the spin-down rate, a change in the pulse profile and a shift in the spectral cut-off to a lower energy, simultaneously around 2011 October 16. To explain these effects on the high-energy emissions by the glitch of PSR~J2021+4026, we hypothesized the glitch to be caused by the rearrangement of the surface magnetic field due to the crustal plate tectonic activities on the pulsar which is triggered by a starquake. In this glitch event, the inclination angle of the magnetic dipole axis is slightly shifted. This proposition is then tested by numerical modeling using a three-dimensional two-layer outer gap model. The simulation results indicate that a modification of the inclination angle can affect the pulse profile and the spectral properties, which can explain the observation changes after the glitch. 
\end{abstract}


\section{Introduction}
Pulsars are fast-spinning neutron stars, with a typical mass of 1.4 - 2.0$M_{\odot}$ and radius of $10^6$cm. The rotational period of a newborn
pulsar ranges from 0.01 to 10s. Pulsars are highly magnetized
with magnetic field strength of the order of $10^{12}$G, and emit electromagnetic radiation
as beams. This beam of radiation can be observed in a pulsation which that has the same period as the rotational period of a pulsar. For an isolated pulsar, the radiation is powered by the rotation. As the rotational energy is eventually consumed, the spin period increases steadily. This spin-down rate is very small. The typical value is $10^{-13}$ s s$^{-1}$ for a young pulsar (e.g. the Crab pulsar), corresponding to an increase of 0.001s after one million years. 

Although every pulsar decelerates through radiation, it is not rare to observe pulsars that experience spin-up. Glitches are internal processes that generate a sudden jump in the rotational period of a pulsar. The exact causes and processes of a glitch are not yet fully understood. Currently, there are two common proposed models: superfluidity and starquake. The superfluid model, first predicted by Packard (1972), suggests that the structure of the pulsar is composed of normal matter and superfluid components. Occasionally, the superfluid component may couple with the normal matter, causing a transfer of angular momentum to the surface of the pulsar, which increases the observed frequency immediately. Another explanation for the glitch is the starquake model, which was first discussed by Ruderman (1969) after the first glitches in the Crab and Vela pulsars were observed. In this model, it is predicted that when a pulsar spins down, the geometrical shape deforms from oblate toward spherical. Since the crust is solid, such deformation accumulates stress in the crust. When the stress is beyond the maximum shear strain that can be supported, the crust cracks and causes a slight change in the shape of the pulsar, and hence the moment of inertia. As the angular momentum is conserved, the pulsar spins up when the moment of inertia is reduced. Espinoza (2011) reported their analysis on 315 glitches observed in the rotation of 102 pulsars. They found that glitches are most extensive in young pulsars with a characteristic age $\tau_c \sim 10$kyr. The Crab pulsar, which is the youngest pulsar in their study, experienced a relatively small glitch size ($\Delta \nu / \nu < 200 \times 10^{-9}$) but the largest $\left| \dot{\nu} \right|$ among the samples. This small glitch size is more likely to be explained by the starquake model, while for pulsars with larger glitch sizes like the Vela pulsar, having $\Delta \nu / \nu > 1000 \times 10^{-9}$, the explanation has to be obtained with the superfluid model. 

In this study, we analyzed the high-energy gamma-ray emission from PSR~J2021+4026. PSR~J2021+4026 is a bright source in gamma rays first discovered by the \textit{Fermi} Gamma-ray Space Telescope (\textit{Fermi}), which is a space observatory launched into a low Earth orbit in 2008. The pulsar is also the first variable gamma-ray seen by the \textit{Fermi} Large Area Telescope (\textit{Fermi}-LAT). There are spectral and timing analyses that indicate the presence of a glitch
in this pulsar. This glitch has several simultaneous effects on the pulsar emissions. We attempt to use a theoretical model and glitch scenario to explain these observed effects. The data analysis of the gamma-ray observations on PSR~J2021+4026 will be presented in Section~\ref{section:observation}. Then, the details of the model used and the model results will be presented in Section~\ref{section:modeling}. Finally, the study is summarized and concluded in Section~\ref{section:conclusion}.

\section{Gamma-ray Observation on PSR~J2021+4026}
\label{section:observation}
\subsection{Overview of PSR~J2021+4026}
PSR~J2021+4026 is a gamma-ray bright pulsar located in the Cygnus region of the Milky Way. Its gamma-ray detection was first discovered by \textit{Fermi}-LAT in the first year of the \textit{Fermi} mission (Abdo et al. 2009a) as 0FGL~J2021.5+4026 with a test-statistic (TS) value of $\sim 4800$, corresponding to a significance of $\sim 70$-$\sigma$, in the energy range from 200 MeV to 100 GeV. The blind frequency search on the LAT data performed by Abdo et al. (2009b) reported that 0FGL~J2021.5+4026 has a rotational period $P=265$ms and a spin-down rate $\dot{P}=5.48 \times 10^{-14}$ s s$^{-1}$. Thus, the source was confirmed to be a pulsar. It has a spin-down characteristic age of $\tau_{c}\sim 77$kyr, an estimated surface magnetic field of $B \sim 4 \times 10^{12}$G and an estimated spin-down luminosity of $\dot{E} \sim 10^{35}$ erg s$^{-1}$. 

A detailed study done by Allafort et al. (2013) on the gamma-ray emission from PSR~J2021+4026 revealed a sudden change in the pulsar behavior that is observed by the \textit{Fermi}-LAT. They found that the glitch happened near MJD 55850 (2011 October 16) with a timescale smaller than a week. The change occurred in four aspects (flux, spin-down rate, pulse profile, and spectrum) simultaneously. First, the flux was decreased by $18\%$, from $(8.33 \pm 0.08) \times 10^{-10}$ erg cm$^{-2}$ s$^{-1}$ to $(6.86 \pm 0.13) \times 10^{-10}$ erg cm$^{-2}$ s$^{-1}$. Second, the magnitude of the spin-down rate was increased from $(7.8 \pm 0.1) \times 10^{-13}$ Hz s$^{-1}$ to $(8.1 \pm 0.1) \times 10^{-13}$ Hz s$^{-1}$ ($\sim 4\%$ increase). Third, the pulse profile was changed significantly ($>5\sigma$). Fourth, the emission spectrum experienced marginal change ($<3\sigma$). In the phase-averaged spectral analysis, the emission from the pulsar is modeled by a power law with exponential cut-off, which has the functional form
\begin{equation}
    \dfrac{\textrm{d}N}{\textrm{d}E}=N_0 \left(\dfrac{E}{E_0}\right)^{-\Gamma} \textrm{exp} \left(-\dfrac{E}{E_C}\right),
    \label{equation:plsec}
\end{equation}
where $N_0$ is the normalization constant, $E_0$ is the scale factor of energy, $\Gamma$ is the spectral power-law index, and $E_C$ is the cut-off energy. It is found that the cut-off energy ($E_C$) decreases by $\sim 13\%$ in the glitch. There is also a slight decrease in the integrated energy flux in the energy range between 100 MeV and 300 GeV. The phase-resolved spectral analysis also revealed that the major variation in the emission spectrum of PSR~J2021+4026 results from the first major peak (P1) of the pulsed signal. The phase-resolved spectrum in the region of P1 experienced a significant drop in the cut-off energy ($E_C$) of $\sim 27\%$ after the glitch. The authors hypothesized that these sudden changes originate from the reconfiguration of magnetic field on the pulsar surface. 

In our study, the gamma-ray emission of PSR~J2021+4026 was re-analyzed, using data of longer length and an updated version of the \textit{Fermi} Science Tools. The phase-averaged and phase-resolved spectral results are consistent with the study by Allafort et al. (2013). From the long-term tracking on the flux emitted from PSR~J2021+4026, we found that the drop caused by the glitch is a permanent effect. In particular, we present the findings from the flux evolution throughout the seven years after the launch of the \textit{Fermi}. 

\subsection{Flux Evolution over Seven Years}
\label{section:dataanalysis}
The gamma-ray flux emitted by PSR~J2021+4026 over seven years since the start of \textit{Fermi} mission is analyzed. The data between 2008 August 04 and 2015 August 12 were used. The events in the Pass 8 “source” class were selected and the corresponding instrumental response functions for this event class is the P8R2\_SOURCE\_V6. The region of interest (ROI) is a $20^\circ \times 20^\circ$ square region centered at the epoch J2000 position $(\textrm{R.A.}, \textrm{Dec})=(20^{\textrm{h}} 21^{\textrm{m}} 34.08^{\textrm{s}},40^\circ 26^\prime 27.6^{\prime\prime})$. All photons with energy between 100 MeV and 100 GeV in this ROI were used. To avoid the contamination from the Earth's albedo, we excluded the time intervals with zenith angles $>90^\circ$ and those with the rocking angle of the LAT being $>52^\circ$. 

The light curve (photon flux against time) of PSR~J2021+4026 was calculated in two steps. In the first step, the background emission model is obtained from the analysis of the data over
seven years. In the second step, the seven-year time interval is binned and the flux from the target pulsar in each time bin is computed. Both steps involve the binned likelihood analysis performed using the \textit{Fermi} Science Tools version v10r0p5\footnote{Available at \url{http://fermi.gsfc.nasa.gov/ssc/data/analysis/software/}}. 

In the first step, the background emission was analyzed using the data over the
entire period ($\sim 7$ yr). The input model contains all 3FGL catalog sources (gll\_psc\_v16.fit; Acero et al. 2015) within $20^\circ$ from the center of ROI. All parameters are fixed
to the 3FGL catalog values for the sources that are $>10^\circ$ away from the ROI center. The galactic diffuse emission (gll\_iem\_v06) and the isotropic diffuse emission (iso\_P8R2\_SOURCE\_V6\_v06), available from the \textit{Fermi} Science Support Center (FSSC)\footnote{\url{http://fermi.gsfc.nasa.gov/ssc/ }}, were also included in the input model. An extended source positioned within the ROI, known as the Cygnus Loop, was modeled with the spatial template (CygnusLoop.fits) provided by the FSSC. Using the gtlike tool in the Science Tools, the best-fit model was obtained. Insignificant sources ($<3$-$\sigma$) were eliminated from the model. All spectral indices in the model were then fixed to the best-fit values. This model is then used as the input model in the second step. 

\begin{figure}
    \centering
    \includegraphics[width=0.9\textwidth]{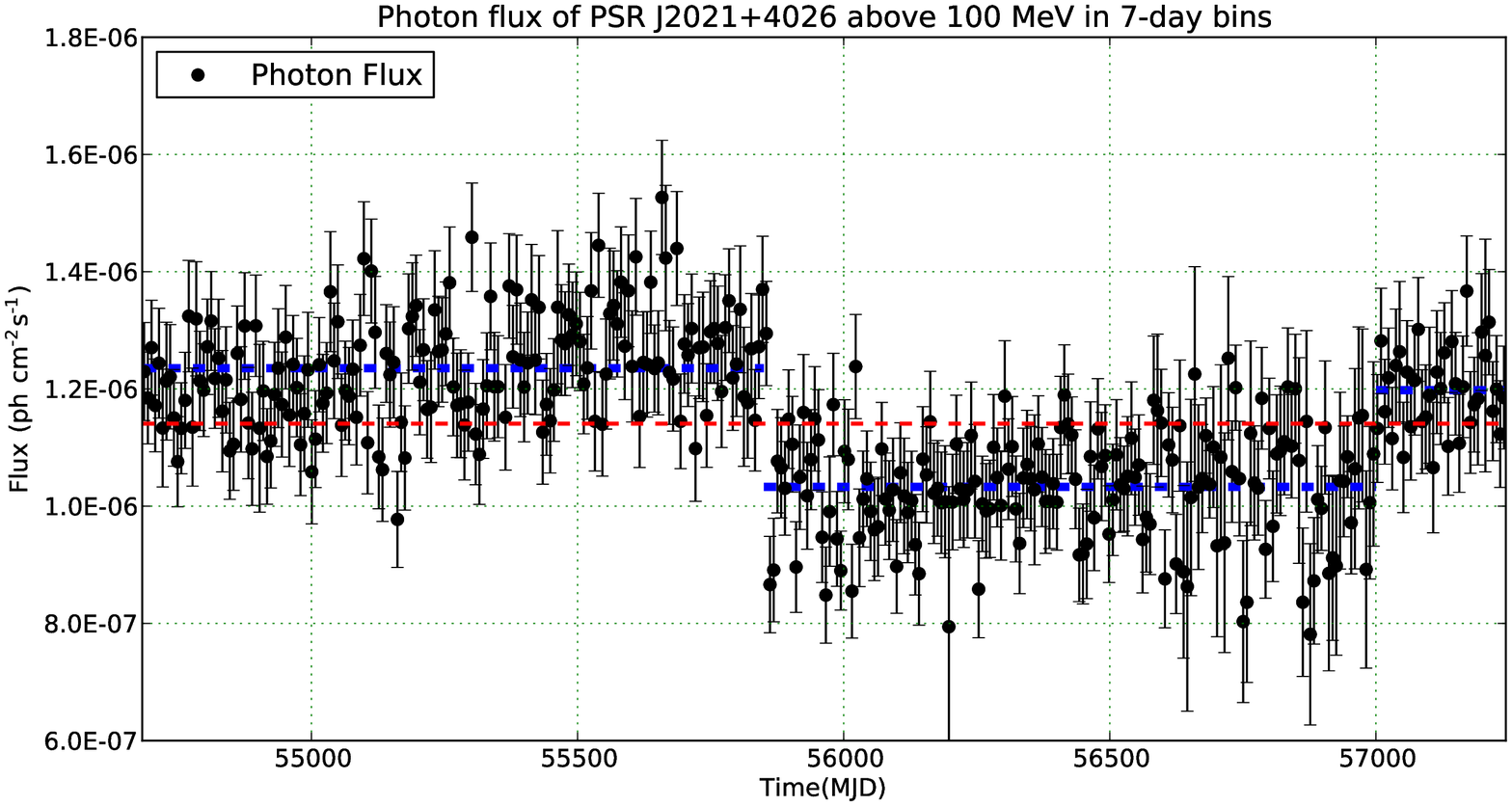}
    \includegraphics[width=0.9\textwidth]{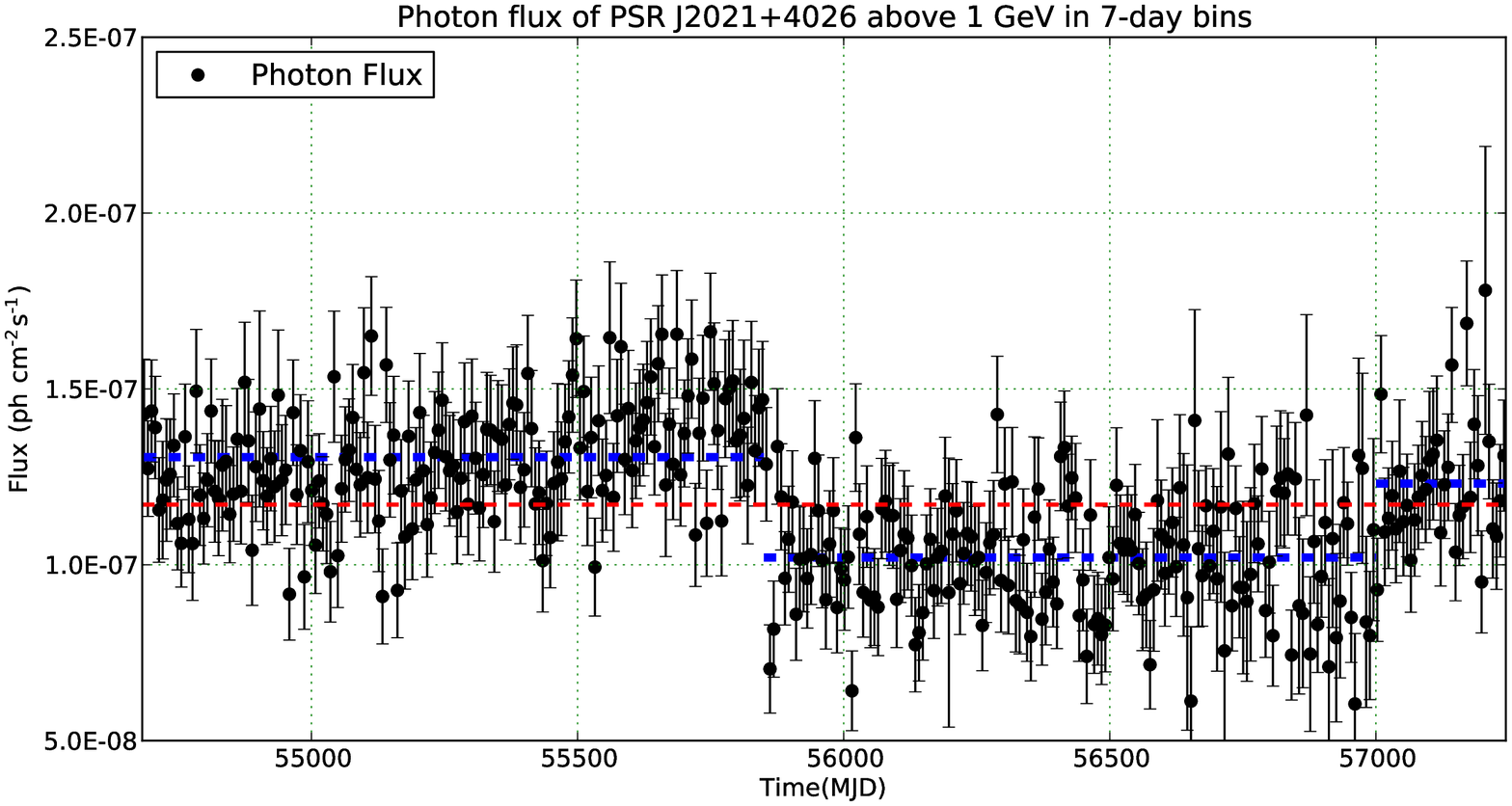}
    \caption{$>100$ MeV (top) and $>1$ GeV (bottom) light curves of PSR~J2021+4026. Seven-year data are binned into bins with an interval of seven days. }
    \label{fig:lc7day}
\end{figure}

\begin{figure}
    \centering
    \includegraphics[width=0.9\textwidth]{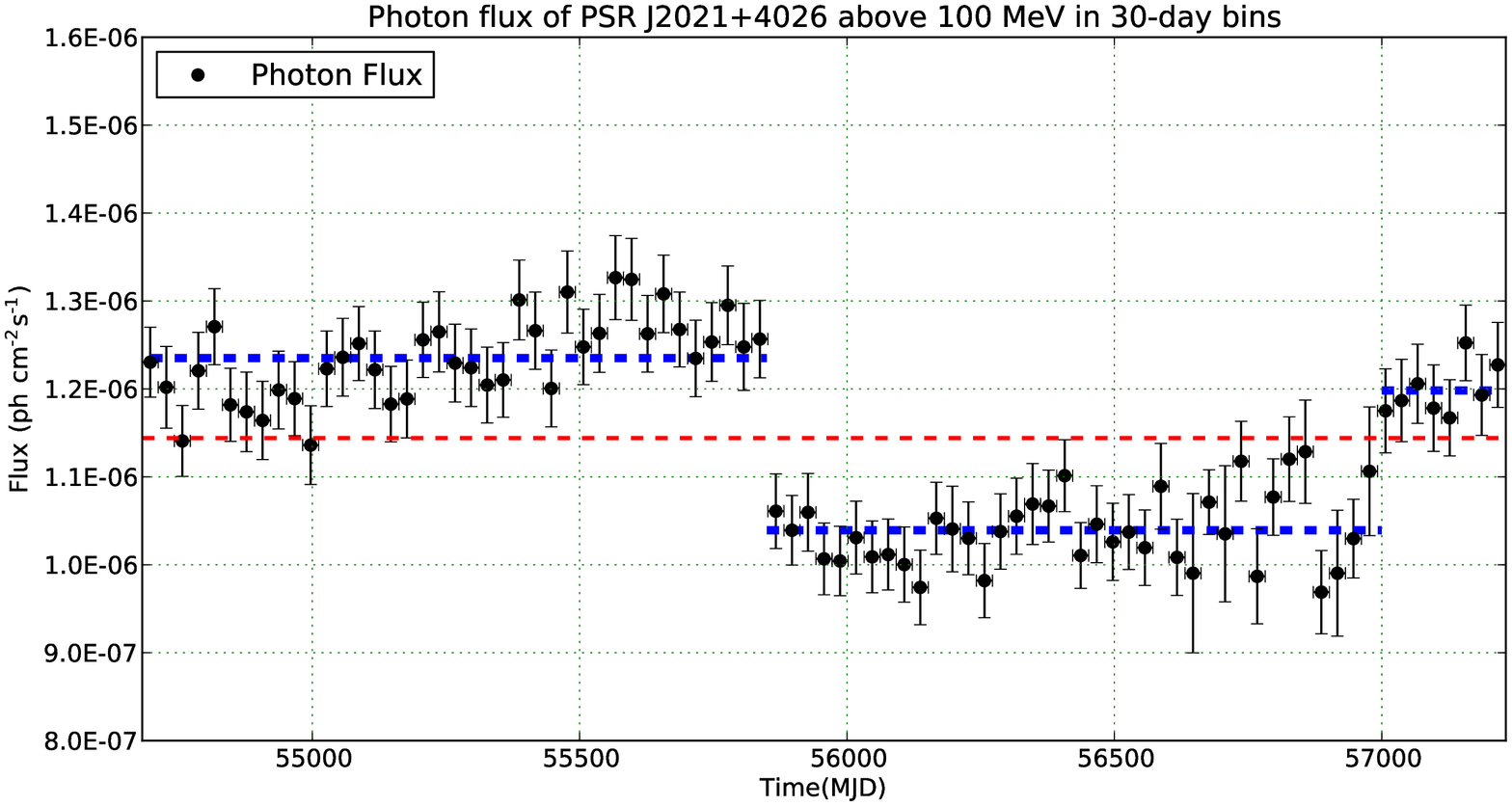}
    \includegraphics[width=0.9\textwidth]{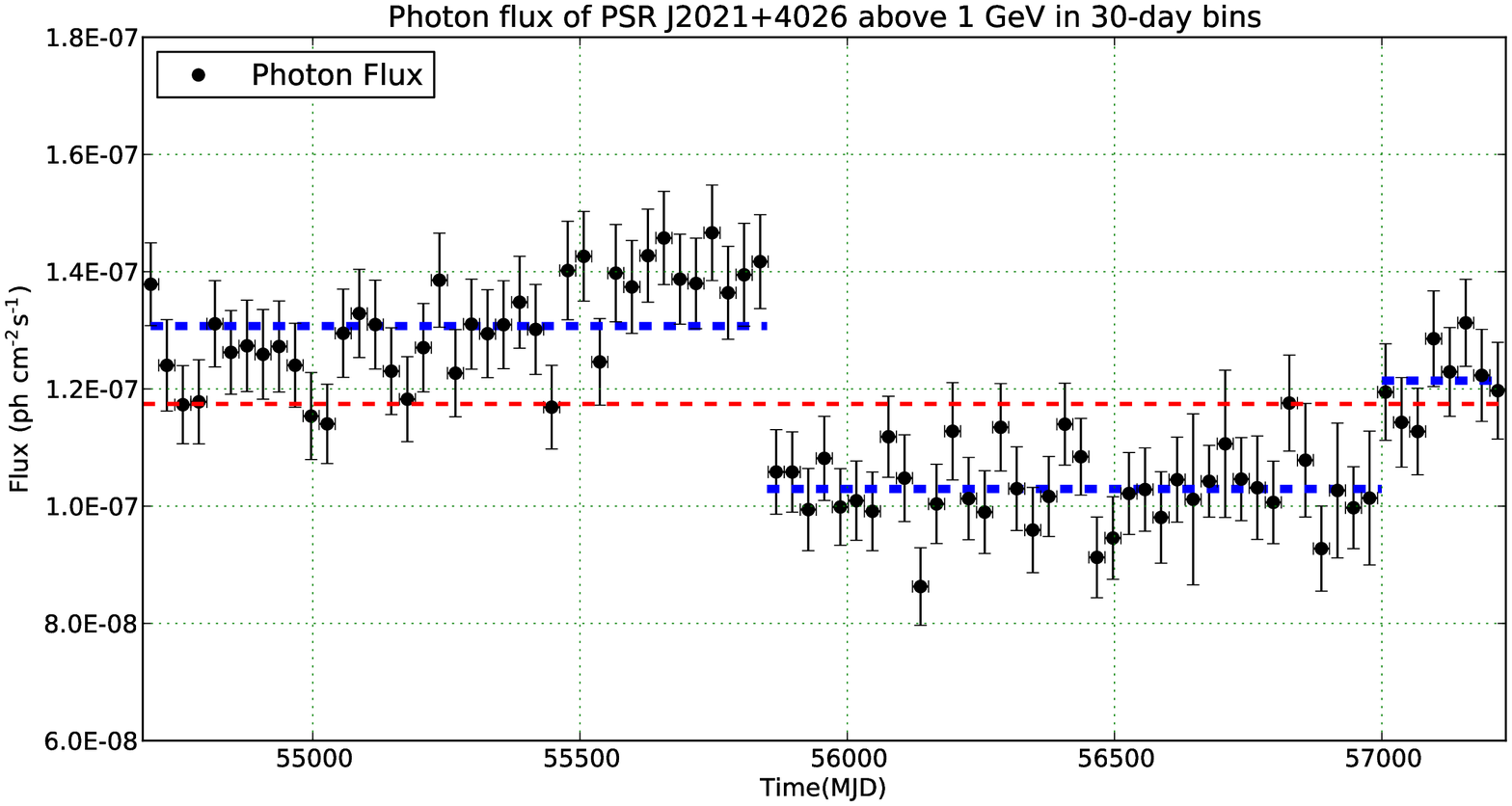}
    \caption{$>100$ MeV (top) and $>1$ GeV (bottom) light curves of PSR~J2021+4026. Seven-year data are binned into bins with an interval of 30 days. }
    \label{fig:lc30day}
\end{figure}

In the second step, the data in seven-years were binned into intervals of 7 days or 30 days.
The photons in two energy ranges, $>100$ MeV and $>1$ GeV, were considered separately. The input model is the best-fit model with fixed spectral indices from the analysis of the entire
seven-years data in the first step. In each time bin, binned likelihood analysis was performed to obtain the normalization constants in the model and to estimate the photon flux of each source included in the model. The photon flux of PSR~J2021+4026 was then obtained. Figure~\ref{fig:lc7day} and figure~\ref{fig:lc30day} show the light curves with energies $>100$ MeV and $>1$ GeV in 7days and 30days bins, respectively. The red dashed line represents the mean flux
value throughout the seven years. The blue dashed lines represent the mean fluxes within three separate intervals: (1) before MJD 55850, (2) between MJD 55850 and 57000, and (3) beyond MJD 57000. As indicated in the light curves, there is a sudden and significant drop ($\sim 20\%$) in the photon flux around MJD 55850. There could also be hints of a second glitch around MJD 57000 in which the flux has experienced another jump with a smaller glitch size. Although we speculate that this could be a second glitch for this pulsar, we cannot confirm it at this moment because we have not obtained the timing parameters beyond MJD 56580. Also, the phase-averaged spectral analysis using data over  eight months, which produces a relatively large uncertainty, does not guarantee a change. Therefore, in this paper, we only discuss the details of the first glitch. As it can be seen from figure~\ref{fig:lc30day}, the flux level between the two jumps is steady and shows no hint of gradual recovery. Therefore, we regarded the first jump in flux as a permanent effect by the glitch in MJD 55850. 

\section{Modeling}
\label{section:modeling}
In this section, we will use a theoretical model to simulate the gamma-ray emissions from PSR~J2021+4026 and explain the cause and the consequences of the glitch that  occurred around 2011 October 16. From the \textit{Fermi}-LAT observations reported in Allafort et al. (2013), the emission of PSR~J2021+4026 exhibits widely separated double peaks connected with a strong bridge, a spectral cut-off at high energy ($\sim 2$ GeV) and a high integrated energy flux. These characteristic properties favor the outer gap model (Cheng et al., 1986a, 1986b). Hence, this study
adopts the outer gap model in  simulating the radiation of PSR~J2021+4026. In particular,
we will use the three-dimensional two-layer outer gap model, which is a more realistic outer gap model that considers the development and the closure of the outer gap by photon-photon pair creation (Wang et al. 2011). The details of the three-dimensional two-layer outer gap model are described in Wang et al. (2010, 2011). We will then use the crust cracking scenario to understand the glitch behavior of PSR~J2021+4026 in Section~\ref{section:crustcrack}. 

\subsection{Glitch of PSR~J2021+4026: the Crust Cracking Scenario}
\label{section:crustcrack}
In this section, the possible cause of the glitch in PSR~J2021+4026 around 2011 October 16 will be discussed. From the gamma-ray observation described in Section~\ref{section:observation},
the 7 years light curve shows that the photon flux of the pulsar experienced a permanent decrease after the glitch. To produce a permanent change, the crust crack mechanism is preferred to the superfluid coupling. This is because the pulsar will gradually recover to its original condition after the glitch caused by superfluid. We propose that the crust cracking scenario leads to a change in the magnetic inclination angle, and thus the observed consequences of the glitch in PSR~J2021+4026. 

\subsubsection{Crust Cracking and the Inclination Angle}
When Link et al. (1992) studied the glitch of the Crab pulsar that occurred in 1975, they found that the superfluid model could not explain the persistent change in the spin-down rate. They proposed that there must be an excess external torque that  is a result of the rearrangement of the surface magnetic field. The surface magnetic field is rearranged through the plate tectonic activity (Ruderman 1991). In the plate tectonic model, the structure of a pulsar consists of an upper crust, a lower crust, and a core. The thickness of the solid crust is $\sim 10^5$ cm. Neutron superfluid vortex lines build up shear stresses in the crust lattices of the spinning pulsar. This stress will break the crust and move the crustal plates toward the equator. Since the surface magnetic field lines are localized and pinned to the plates, they move along with the plates. As a result, the dipole moment of the pulsar is more deviated from the rotational axis, thus increasing the magnetic inclination angle. In the case study of the  Crab pulsar by Link et al. (1992), they found that a change in the inclination angle $\Delta \alpha \sim 10^{-4} \tan \alpha$ could reproduce the observed offset in the spin-down rate $\Delta \dot{\Omega}$. 

\subsubsection{Effect on Spin-down Rate}
\label{subsection:spindownangle}
One main approach to describe the magnetosphere of pulsars is by solving equations for magnetohydrodynamics (MHD) in the force-free limit, where the forces in the plasma are balanced (i.e. $\rho \vec{E}+\frac{1}{c}\vec{j}\times\vec{B}=0$, where $\rho$ and $\vec{j}$ are the charge and current densities, respectively). The equations for MHD are derived from the Maxwell equations in special relativity with the imposed force-free conditions. Spitkovsky (2006) solved the MHD equations numerically using a finite-difference time-domain (FDTD) approach to investigate the evolution of pulsar magnetospheres. From the numerical solutions, it is found that for a pulsar having an inclination angle between the magnetic and the rotational axes, the spin-down luminosity can be formulated by
\begin{equation}
L_{pulsar}=k_1\dfrac{\mu^2 \Omega^4}{c^3}(1+k_2\sin^2\alpha),
\label{equation:spindownluminosity}
\end{equation}
where $k_1=1\pm 0.05$ and $k_2=1\pm 0.1$ are the best-fit coefficients from the numerical results. Considering the outer gap model as a small perturbation of the force-free model, we adopted this relationship. Using $L_{pulsar}=I\Omega\dot{\Omega}$, where $I$ is the moment of pulsar's
inertia, we found that the relative change in the spin-down rate ($\Delta\dot{\Omega}/\dot{\Omega}$) can be expressed as a function of the inclination angle ($\alpha$):
\begin{equation}
\dfrac{\Delta\dot{\Omega}}{\dot{\Omega}}=\dfrac{\sin 2\alpha\Delta\alpha}{1+\sin^2\alpha}.
\label{equation:sdinclination}
\end{equation}
In this way, a shifting in the inclination angle can lead to a change in the spin-down rate of the pulsar. In the case of PSR~J2021+4026, since the spin-down rate is increased after the
glitch, we expect there to be  an increase in the inclination angle. 

\subsubsection{Effect on Pulse Profile and Cut-off Energy}
Because of the rearrangement of surface magnetic fields due to crust plate activities, it is expected that the emission geometry is affected. The observed pulse profile can be interpreted as the pattern of magnetic field lines viewed at a fixed viewing angle $\beta$, which is the angle between the rotational axis of the pulsar and the observer. Higher flux is observed at the positions with higher magnetic field line densities. This is because the density of the magnetic
field line corresponds to the field strength. If the magnetic field is stronger, the charged particles are accelerated to higher speeds and emit more energetic curvature photons. The secondary pair creation and acceleration also contribute to the high flux observed. When the inclination angle is changed, leading to an adjustment in the magnetic field line pattern, the observed pulse profile will be modified. 

Another consequence is the shift in the spectral cut-off energy. As discussed in Section~\ref{subsection:spindownangle}, the inclination angle of PSR~J2021+4026 is expected to increase after the glitch. When the inclination angle increases, the line of sight from the observer will cut the magnetic field lines at an angle closer to the polar region. Therefore, the inner magnetosphere will be viewed. Since the charged particles are accelerated through the outer gap toward the light cylinder, the curvature photons from the inner part have less energy than the outer region. As a result, the observed emission with increased inclination angle generally has lower energy. Thus, the spectral cut-off energy is shifted to a smaller value. 

\subsection{Results}
We used the three-dimensional two-layer outer gap model described in Wang et al. (2011) to simulate the high-energy emission from PSR~J2021+4026. It is found that the model could not reproduce the widely separated peaks and high energy flux shown in the observation simultaneously. This is because a larger peak separation requires a thinner gap but a thinner gap will result in a lower spectral energy flux. To preserve both properties, we introduced an extra parameter to increase the energy in the emission spectrum such that a thinner gap can be applied. Takata \& Chang (2009) modeled the dependency between the maximum accelerating electric field in the outer gap structure and the size of the polar cap. It is found that the typical strength of the accelerating electric field is proportional to the square of the polar cap size. Here, we used an enlargement factor to amplify the size of the polar cap. In this way, both the peak separation and the high-energy emission spectrum can be retained. We chose the factor to be $1.6$. 

The observation on PSR~J2021+4026 shows that the pulsar releases strong off-pulse emission, which is difficult to explain by traditional outer gap model. In order to produce the off-pulse emission, the outer gap is extended towards the stellar surface, until a limit called the inner boundary (Dyks \& Rudak 2003; Dyks et al. 2004 and Takata et al. 2004). Also, the viewing angle is required to be close to $90^\circ$ in order to observe the off-pulse emissions. In this study, the inner boundary was chosen to be located at the height of $(1-f)$ times the distance between the stellar surface and the null charge surface, where $f$ is the outer gap fraction defined in Wang et al. (2011). 

In this section, the results will be presented. In particular, the pulse profile  (in the full energy band and in separate energy bins) and spectra (phase-averaged and phase-resolved) before and after the glitch will be discussed. 

\subsubsection{Pulse Profiles and Phase-averaged Spectra}
In the model, the pulsar PSR~J2021+4026 has a surface magnetic field of $1 \times 10^{13}$G at the poles, a rotational period of $0.265$~s
and a distance of $1.0$ kpc away from the Earth. The model parameters that characterize the geometry of the pulsar and the distributions of the properties of the two-layer outer gap are presented in table~\ref{table:modelpara}. $\alpha$ and $\beta$ are the inclination angle and viewing angle,  respectively. $C$ is the normalization constant for the distribution of the gap fraction $f$. $B_1$ and $B_2$ are the normalization constants for the distribution of the ratio, $h_1/h_2$, between the sizes of the main acceleration ($h_1$) and screening ($h_2$) regions. $F$ is the normalization parameter for the drift motion of charged particles. $A$ is the offset parameter for the distribution of the average charge density. The expressions of these distributions are described in Wang et al. (2011). $\alpha$ and $\beta$ are determined from the pulse profile that
resembles the geometry of the magnetic field structure of a rotating dipole. Other parameters characterize the strength of the accelerating electric potential in the outer gap and thus the energy of the radiated curvature photons. 

To test the crust cracking scenario on this pulsar glitch, we first obtained the set of parameters based on the observed pulse profile and spectrum. Then, we increased the value of the inclination angle $\alpha$ until there is no bump in the bridge emission of the pulse. The value of the viewing angle $\beta$ does not change after glitch because it is the angle between the rotational axis and the observer, which is fixed with time. Since we intend to explain the glitch by the change in the inclination angle, other parameters involved in the distributions of the gap structure are assumed to be unchanged to reduce the number of free parameters and the complexity of the scenario. From the relation between the change in the spin-down rate and the inclination angle discussed in Section~\ref{subsection:spindownangle}, we can estimate $\Delta \alpha$ from the observed $\Delta \dot{\Omega}$. Using equation~\ref{equation:sdinclination}, the observed value of $\dfrac{\Delta \dot{\Omega}}{\dot{\Omega}} \sim 4\%$ and the modeled value of $\alpha \sim 63^\circ$ before the glitch, we obtained the expected change in the inclination angle during the glitch to be $\Delta \alpha \sim 3^\circ$. This expectation of a $\sim 3^\circ$ increase in the inclination angle is consistent with the modeled value of $4.2^\circ \pm 0.3^\circ$, which is the difference between the values before and after glitch listed in table~\ref{table:modelpara}. 

\begin{figure}
    \centering
    \includegraphics[width=1.\textwidth]{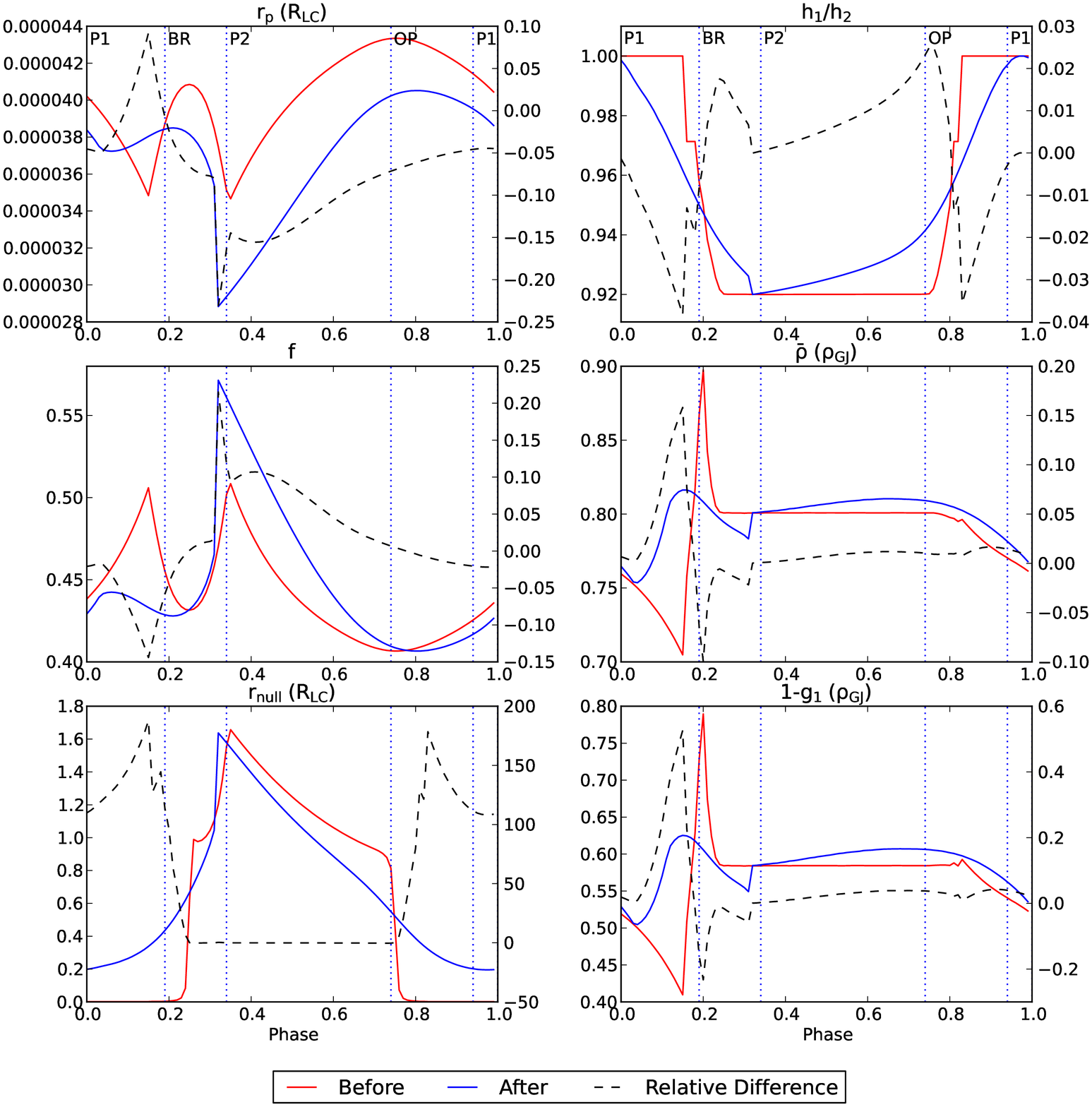}
    \caption{Two-layer structural azimuthal distributions of (top left) polar cap radius, (top right) ratio between sizes of two layers, (middle left) gap fraction, (middle right) average charge density, (bottom left) radial distance to null charge surface surface,  and (bottom right) charge density in the primary acceleration region. Red and blue solid lines represent the distributions before and after glitch, respectively. Black dashed lines are the relative changes in the distributions after the glitch. The y-axis on the left of each plot indicates the values of the red and blue curves. The y-axis on the right of each plot indicates the value of the black curve. }
    \label{fig:distdiff}
\end{figure}

\begin{figure}
    \centering
    \begin{tabular}{cc}
    	\includegraphics[width=0.45\textwidth,height=0.65\textheight]{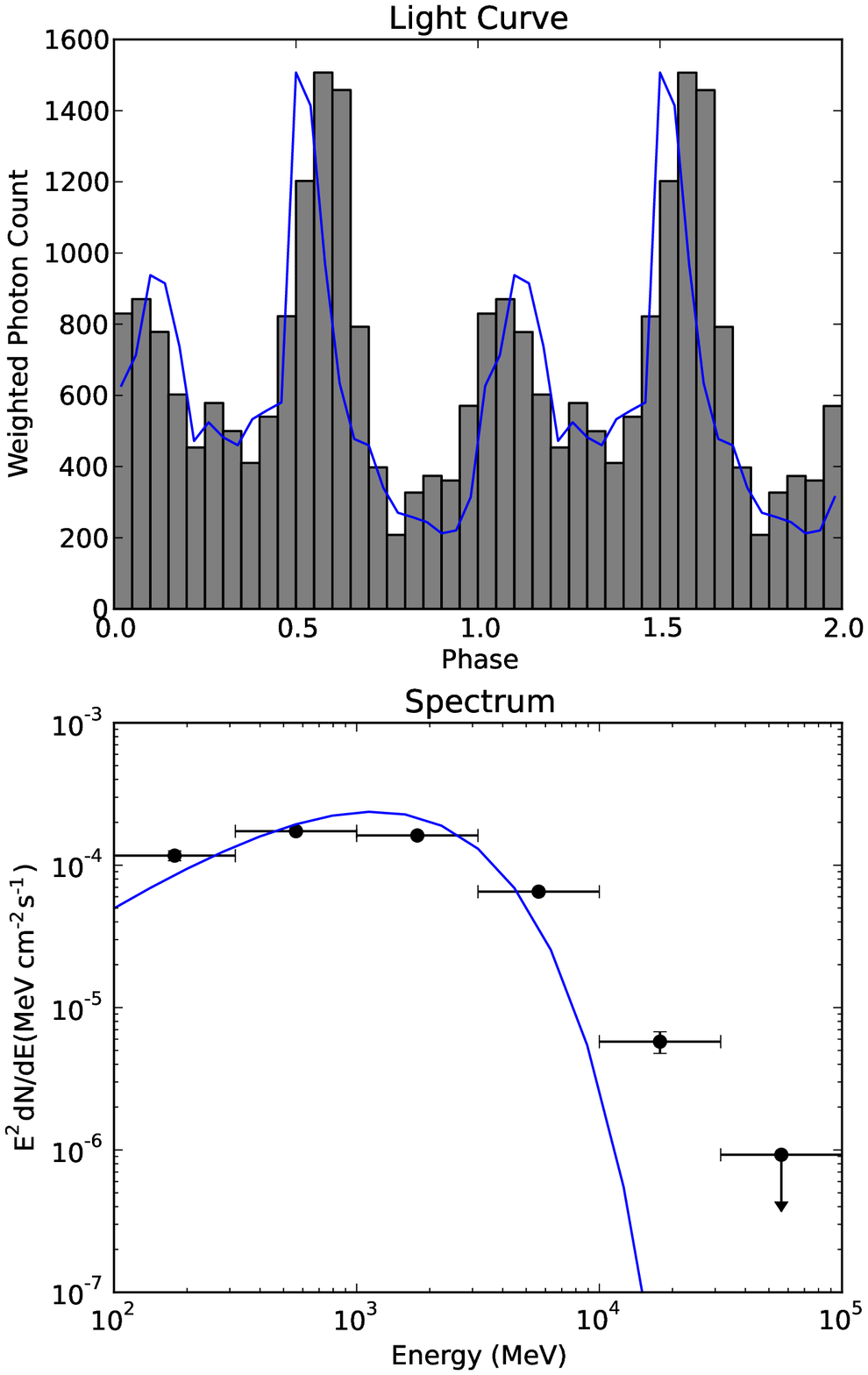} &
    	\includegraphics[width=0.45\textwidth,height=0.65\textheight]{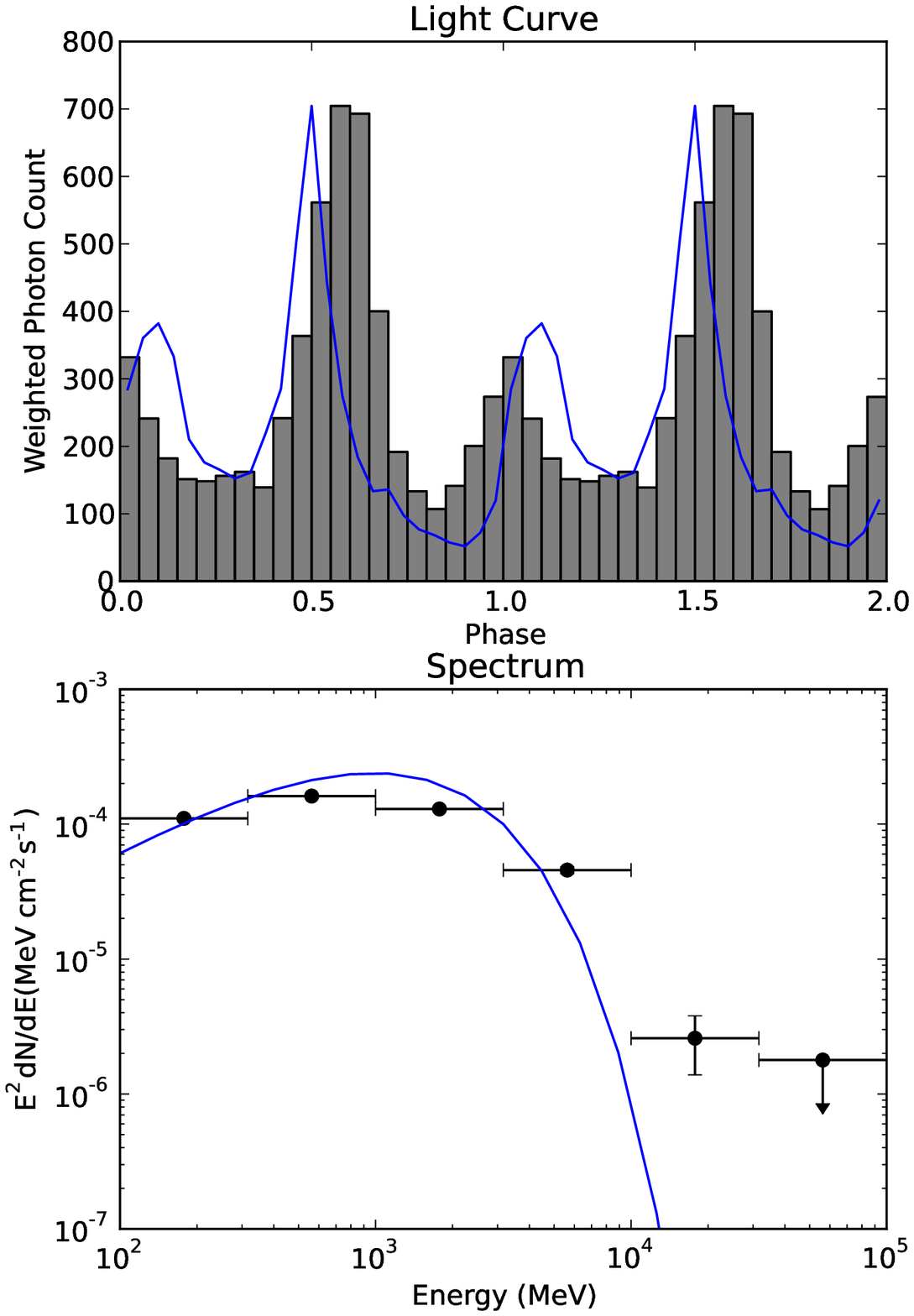} \\
    \end{tabular}
    \caption{The pulse profile (top) and phase-averaged spectra (bottom) for PSR~J2021+4026 before (left) and after glitch (right). Blue solid lines are the model pulse profile and spectrum. Histogram is the observed pulse profile from \textit{Fermi}-LAT data. Normalization is applied to match the intensities of the modeled and data pulse profiles. Black dots with errors are the SEDs estimated from the analysis of \textit{Fermi}-LAT data. }
    \label{fig:lcspecbeforeafter}
\end{figure}

The model results are shown in figure~\ref{fig:distdiff}-\ref{fig:lcspecbeforeafter}. The structural distributions in the outer gap before and after the  glitch are shown
in figure~\ref{fig:distdiff}. The plots include the distributions of the polar cap
radius $r_p$, the  gap fraction $f$, the size ratio of the two layers $h_1/h_2$,
the radial distance to null charge surface $r_{null}$, the average charge density $\bar{\rho}$,
and the charge density in the acceleration region. The red and blue solid lines represent the distributions before and after the glitch, respectively. The black dashed lines are the
relative changes in the distributions after the glitch calculated by $(x_f-x_i)/x_i$, where $x$ is the concerned quantity. 

The model pulse profile and spectrum before the glitch is shown in the left panel of figure~\ref{fig:lcspecbeforeafter}. The model pulse profile is drawn by a blue solid line and the observed pulse profile is represented by a gray histogram for comparison. The observed pulse profile is a photon-weighted folded curve using the data events between 2008 August 11 and 2011 December 16
on an ROI of $1^\circ$ centered at PSR~J2021+4026. The pulsar ephemeris for PSR~J2021+4026 is obtained from the \textit{Fermi}-LAT Multiwavelength Coordinating Group \footnote{\url{https://confluence.slac.stanford.edu/display/GLAMCOG/LAT+Gamma-ray+Pulsar+Timing+Models}} (Ray et al. 2011). The weighting was done by the gtsrcprob tool in the Science Tools, which calculates the probability that a photon originates from the pulsar, based on the maximum likelihood model in the same time period. The model was computed by the same method as described in Section~\ref{section:dataanalysis}. As seen in the pulse profile, the bump between the double peaks is a result of the outer gap geometry in the model, which states there is a local maximum in the magnetic field line density between the two peak emission regions. The peak separation and the off-pulse emission are also generally consistent with the observed data. The emission spectrum before the glitch is shown in the bottom left panel of figure~\ref{fig:lcspecbeforeafter}. The data points are from the spectral analysis in the same period of time as the pulse profile. An upper limit is reported when the detection significance is less than $3\sigma$.

The right panel in figure~\ref{fig:lcspecbeforeafter} shows the model pulse profile and spectrum after the  glitch. The observed pulse profile uses the data between 2011 December 16 and 2013
October 14 limited by the valid time range of the timing model. In the case of post-glitch, there is no bump between the double peaks in the model. This corresponds to the vanish of the bump as seen in the gamma-ray observation. The cause of the bump disappearance is mainly the change of the inclination angle $\alpha$. The off-pulse emission is generally consistent with the observed data. However, the peak separation is less comparable to the data than in the case of pre-glitch. This is due to the fact that we used the inclination angle $\alpha$ as the only parameter in this simple model. $\alpha$ affects the geometry of the magnetic field lines. When it is increased, the inner layer of the magnetosphere is observed at the line of sight. Therefore, the span of the outer gap emission during a pulse period is shorter and the peak separation becomes smaller. For the spectrum, the energy flux at energy $>10$GeV is not well modeled and is underestimated when compared to the observation. It is the model's limitation to produce $>10$ GeV photons because the model describes a static outer gap thickness (Takata et al. 2016). 

\subsubsection{Energy-dependent Pulse Profiles and Phase-resolved Spectra}
\begin{figure}
    \centering
    \begin{tabular}{cc}
    \includegraphics[width=0.4\textwidth]{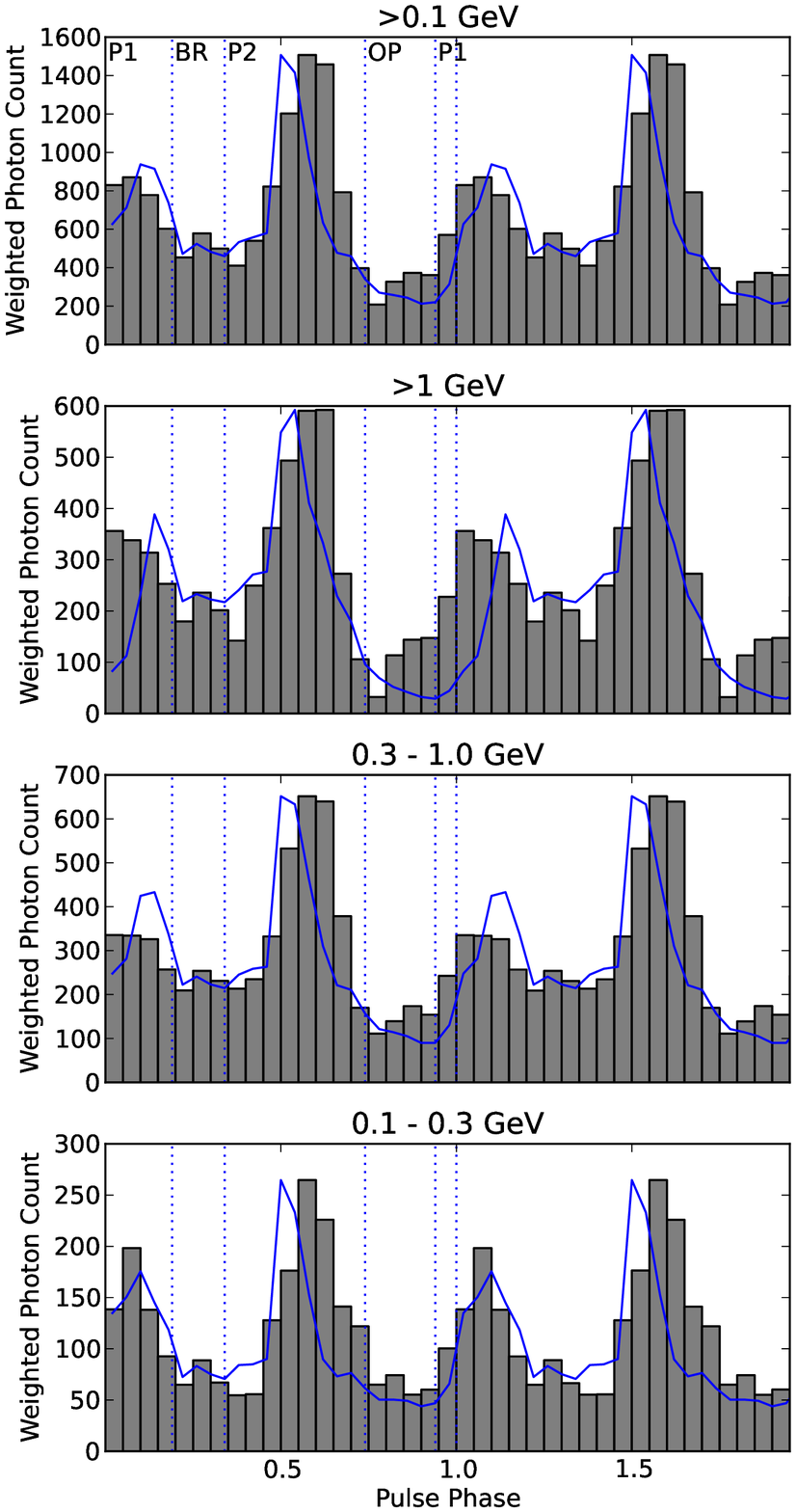} &
    \includegraphics[width=0.4\textwidth]{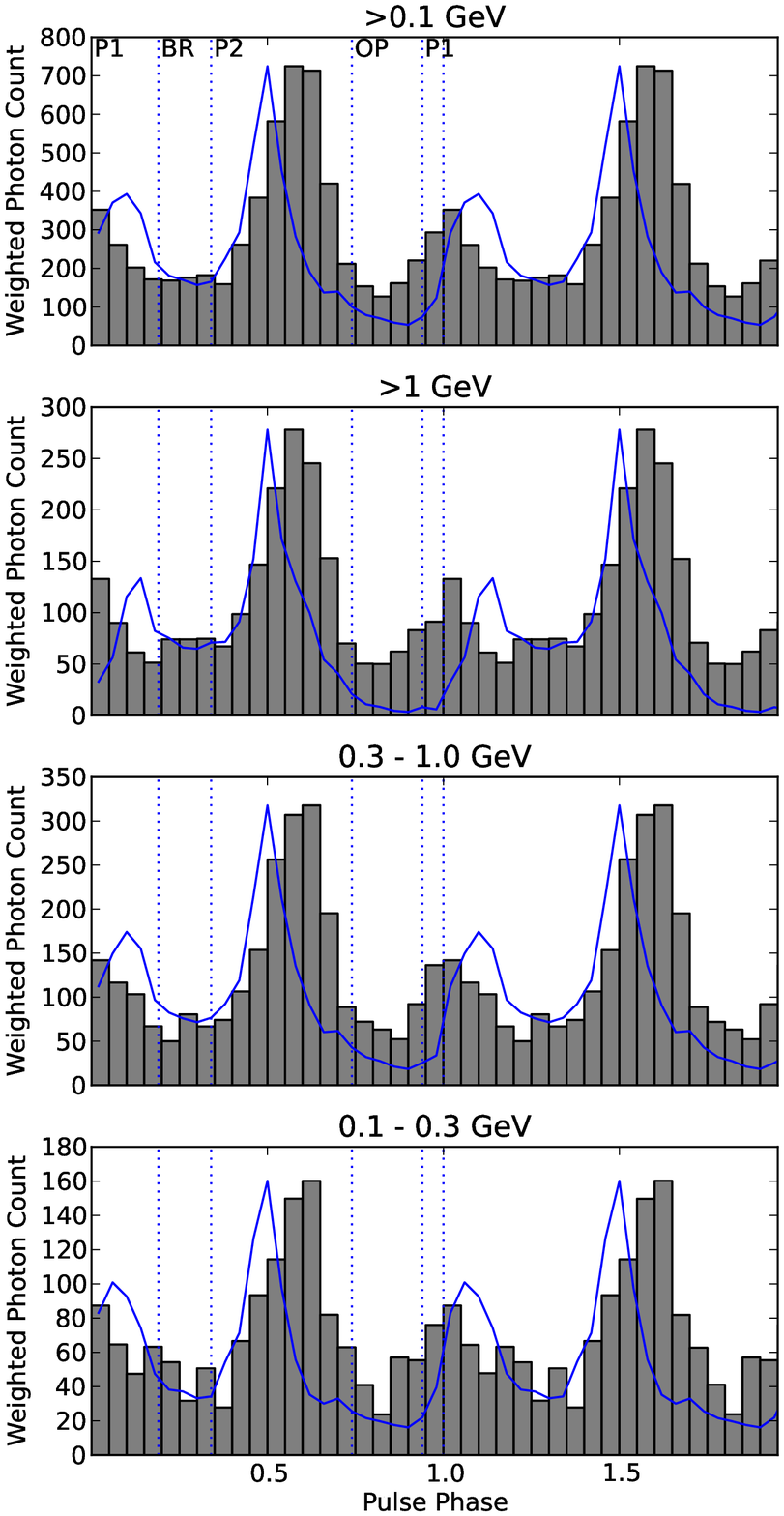} \\
    \end{tabular}
    \caption{Energy-dependent pulse profiles of PSR~J2021+4026 before (left) and after the glitch (right). From top to bottom, the energy ranges are $>0.1$ GeV, $>1$ GeV, $0.3-1.0$ GeV, and $0.1-0.3$ GeV. The blue solid line is the model pulse profile. The histogram is the observed pulse profile from \textit{Fermi}-LAT data. Normalization is applied to match the intensities of the modeled and data pulse profiles. }
    \label{fig:edlcbeforeafter}
\end{figure}

Through selecting different energy ranges for integrating the flux, the energy-dependent pulse profile is obtained. Figure~\ref{fig:edlcbeforeafter} illustrates the energy-dependent pulse profile of PSR~J2021+4026 before and after the glitch, respectively. The same energy bands used in data analysis were applied to the model pulse profile: (1) $>0.1$ GeV, (2) $>1$ GeV, (3) $0.3-1.0$ GeV,  and (4) $0.1-0.3$ GeV. 

\begin{figure}
    \centering
    \hspace*{-0.1\textwidth}\includegraphics[width=1.2\textwidth]{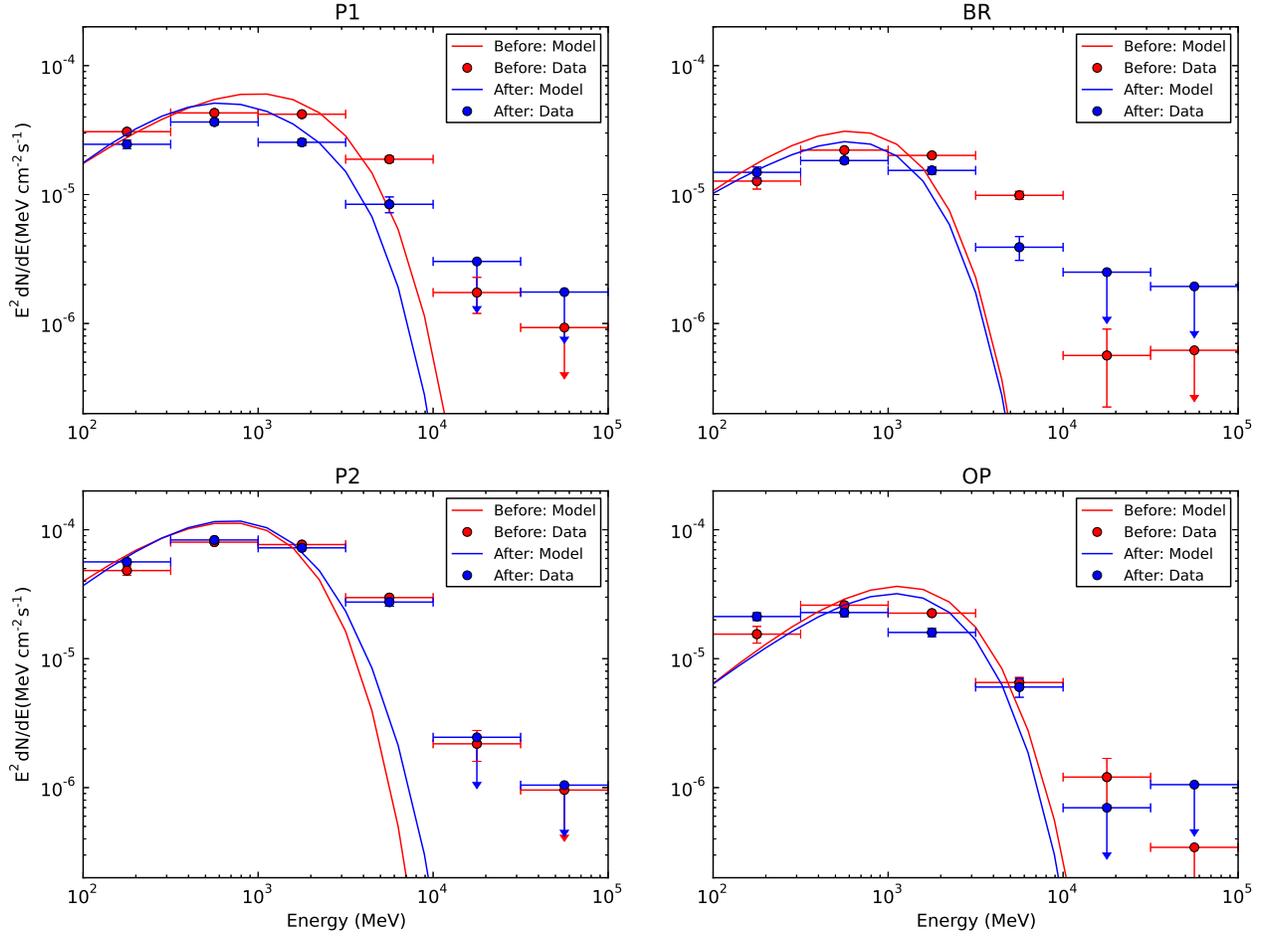}
    \caption{Phase-resolved spectra of PSR~J2021+4026. Red and blue solid lines represent the spectra calculated from the three-dimensional two-layer outer gap model before and after the
      glitch, respectively. Red and blue dots are the SEDs estimated from  the \textit{Fermi}-LAT gamma-ray data before and after the glitch, respectively. The four phases are (top left) first major peak (P1), (top right) bridge (BR), (bottom left) second major peak (P2), and (bottom right) off-peak (OP). }
    \label{fig:phaseresolvedspec}
\end{figure}

The phase-resolved spectra are obtained by selecting the pulse intervals: first major peak (P1), bridge (BR), second major peak (P2), and off-pulse emission (OP), as labeled in figure~\ref{fig:edlcbeforeafter}. The results are shown in figure~\ref{fig:phaseresolvedspec}. Both spectra before and after glitch are plotted on the same figure with the red line representing the results before the glitch and the blue line representing the results after glitch. As discussed in Section~\ref{section:observation}, the decrease in the cut-off energy in the spectrum of phase P1 is much more significant among the phase-resolved spectra. Here, the three-dimensional two-layer outer gap model reproduced this special feature. We can see that the cut-off energy in the P1 spectrum in figure~\ref{fig:phaseresolvedspec} experienced a decrease after the glitch, while in other phases, the spectral cut-off energies are approximately maintained at the same values. The current model did not give a well-fit spectrum for the bridge emission. This is because the phase interval for BR is the narrowest among the four stages so that errors are accumulated most easily. 

The sharp decrease of the spectral cut-off energy in P1 can be understood by considering the electrical properties in the two layers during the glitch. The relative changes of the azimuthal distributions for these properties of PSR~J2021+4026 are plotted in figure~\ref{fig:distdiff}. Phase intervals (P1, BR, P2, and OP) are indicated on the figure. The gap fraction $f$ shows a $\sim 15\%$ decline during P1 after the glitch. A smaller gap fraction means a smaller outer gap
region. The decrease in the electric potential difference across the gap establishes a weaker electric field in the acceleration region. The charged particles are accelerated to a slower speed and emit less energetic curvature photons, which will eventually suppress the cut-off energy. In another view, the ratio between the sizes of the two layers $h_1/h_2$ in the outer gap also shows a $\sim 4\%$ decrease in P1. This means that the ratio between the amount of $\sim$ GeV and $\sim 10^2$ MeV photons emitted in the acceleration and screen regions is lowered. Assuming
that the amount of GeV photons is subject to a heavier reduction than the amount of hundred MeV photons, the resulting drop in the cut-off energy shown in the phase-resolved spectrum of P1 can be explained. 

On the other hand, the spectral cut-off in P2 shows no significant change after the glitch.
This can also be explained by the gap structure as indicated by figure~\ref{fig:distdiff}. The gap fraction $f$ and  the ratio $h_1/h_2$ are increased by $\sim 5\%$
and $\sim 1\%$, respectively. The magnitude of these changes is much less than those that occurred in P1. Indeed, the energy flux of P2 at higher energy $(>5\rm{GeV})$  is slightly enhanced. 

\section{Conclusion}
\label{section:conclusion}
We re-analyzed the \textit{Fermi}-LAT data by using a data length of seven years and an energy range from $100$ MeV to $100$ GeV. It is found that the flux of PSR~J2021+4026 experienced two jumps separated by more than three years. In the first jump, the flux was decreased by $\sim 20\%$ and it shows no hint of recovery. Therefore, we regarded the flux change as a permanent effect of the glitch. Although we speculate the second jump to be caused by another glitch, it cannot be confirmed at this moment. Concerning the first glitch, we proposed the crust cracking scenario to explain the glitch event, in which we assumed that the glitch was caused by plate tectonic activity. The magnetic field lines frozen on the surfaces were carried by the movement of these plates, resulting in the change of the magnetic inclination angle. This variation in the inclination angle will have an effect on the spin-down rate of the pulsar, the pulse pattern as seen by the observer, and the characteristic energy of the photons viewed at a constant line of sight. 

We modeled the pulse profile and spectra of PSR~J2021+4026 using the three-dimensional two-layer outer gap model and simulate the glitch effect by imposing an increase in the inclination angle. In the model parameters, the inclination angle is increased by $\sim 4^\circ$, which is consistent with what we expect from the relation to the observed increase in the spin-down rate. Other parameters are fixed for the purpose of model simplification and control. We generated the pulse profiles, in both energy integrated and energy dependent, and the spectra, in both phase-averaged and phase-resolved, for PSR~J2021+4026. We found from the pulse profiles that the bump disappearance can be explained. We also found that the spectral cut-off is decreased most significantly in the pulse phase interval P1. We argued from the azimuthal distributions of the electrical properties in the interval of P1 that the size of the outer gap in the region of P1 is significantly reduced after the glitch, leading to a weaker electrical acceleration of the charged particles. Moreover, the ratio between the sizes of the acceleration and the screening regions is also reduced, which results in the relatively greater reduction in the high-energy photons at GeV. These conditions together reproduced the observed drop in the spectral cut-off energy. 

We express our appreciation to an anonymous referee for useful comments
and suggestions. We thank  A. H. Kong, C. Y. Hui, P. H. T. Tam, M. Ruderman,
and S. Shibata for the useful discussions. C.W.Ng N.G. and K.S.Cheng
are supported by a GRF grant from  the  Hong Kong Government under HKU17300814P. J.Takata is supported by the NSFC grants of China under 11573010.

\begin{table}
    \renewcommand*{\arraystretch}{1.4}
    \centering
    \begin{tabular}{lcc}
        \hline
        Parameters & Before & After \\
        \hline
        \hline
        $\alpha$ & $63.2 \pm 0.1$ & $67.4 \pm 0.2$ \\
        $\beta$ & \multicolumn{2}{c}{94.2} \\
        $C$ & \multicolumn{2}{c}{0.4066} \\
        $B_1$ & \multicolumn{2}{c}{0.92} \\
        $B_2$ & \multicolumn{2}{c}{0.08} \\
        $F$ & \multicolumn{2}{c}{-45} \\
        $A$ & \multicolumn{2}{c}{$1 \times 10^{-3}$} \\
       \hline
    \end{tabular}
    \caption{Values of the three-dimensional two-layer outer gap model parameters described
      in Section~\ref{section:modeling}. $\alpha$ and $\beta$ are the inclination angle and viewing angle, respectively. $C$ is the normalization constant for the distribution of the gap fraction $f$. $B_1$ and $B_2$ are the normalization and offset constants for the distribution of the ratio between sizes of the acceleration and screening regions $h_1/h_2$. $F$ is the normalization parameter for the drift motion of charged particles. $A$ is the offset parameter for the distribution of the average charge density. In order to test the proposed crust cracking scenario, only the inclination angle is subject to change. Other parameters are fixed to simplify the situation. 
    \label{table:modelpara}}
\end{table}
\end{document}